\documentstyle[12pt]{article}
\def\beq{\begin{equation}}
\def\eeq{\end{equation}}
\def\bea{\begin{eqnarray}}
\def\eea{\end{eqnarray}}

\def\ba{\begin{array}}
\def\ea{\end{array}}
\def\d{\partial}

\begin{document} 
\baselineskip17pt
\begin{center} 
{\large \bf \sf
Fractional statistics in some exactly solvable \\ Calogero-like 
             models with PT invariant interactions }

\vspace{1.5cm}

{\sf B. Basu-Mallick},
\footnote{E-mail address~: biru@theory.saha.ernet.in} 

\bigskip

{\em  Saha Institute of Nuclear Physics,  Theory Group \\
 1/AF Bidhan Nagar, Kolkata 700 064, India }

\bigskip

\end{center} 
\vspace {1.75 cm} 
\baselineskip=15.5pt 
\noindent {\bf Abstract }

Here we review a method for constructing exact eigenvalues and eigenfunctions
of a many-particle quantum system, which is obtained by adding some 
nonhermitian but PT invariant (i.e., combined parity and time reversal 
invariant) interaction to the Calogero model. It is shown that such extended 
Calogero model leads to a real spectrum obeying generalised exclusion 
statistics. It is also found that the corresponding exchange statistics 
parameter differs from the exclusion statistics parameter and exhibits a 
`reflection symmetry' provided the strength of the PT invariant interaction 
exceeds a critical value. 

\vspace {.05 cm} 

\newpage 
 
\noindent \section {Introduction }
\renewcommand{\theequation}{1.{\arabic{equation}}}
\setcounter{equation}{0}

\medskip

It is well known that integrable dynamical models and spin chains with 
long range interactions exhibit fractional statistics or generalised exclusion
statistics (GES) \cite{ha}, 
which is believed to play an important role in many strongly correlated
systems of condensed matter physics.
The $A_{N-1}$ Calogero model (related to $A_{N-1}$ Lie algebra)
is the simplest example of such dynamical model,  containing $N$ 
particles on a line and with Hamiltonian given by
\cite {ca,su}
\beq
H= -  {1\over 2} \sum_{j=1}^N {\d^2 \over \d x_j^2}
+   {\omega^2\over 2}  \sum_{j=1}^N x_j^2 +
   {g \over 2} \sum_{j\neq k} {1 \over (x_j -x_k)^2} \, , 
\label{1} 
\eeq
where $g$ is the coupling constant associated with long-range interaction. 
One can exactly solve this Calogero model and find out the complete set of
energy eigenvalues as 
\beq 
E_{n_1,n_2, \cdots , n_N} = 
 {N \omega \over 2} \left [ 1 + (N-1)\nu \right ] + \omega \sum_{j=1}^N n_j .
\label {2}
\eeq
Here $n_j$s are non-negative integer valued 
quantum numbers with $n_j \leq n_{j+1}$ and $\nu $
is a real positive parameter which is related to $g$ as 
\beq 
g = \nu^2 - \nu \, .
\label {3}
\eeq
It may be noted that, apart from a constant shift for all energy levels, 
the spectrum (\ref {2}) coincides with that of $N$ number of free 
bosonic oscillators. Furthermore,  one can easily  remove the above mentioned 
 constant shift for all energy levels and express (\ref{2}) 
 exactly in the form of energy eigenvalues for free oscillators:  
$E_{n_1,n_2, \cdots , n_N} = {N \omega \over 2}  + \omega 
\sum_{j=1}^N \bar n_j, 
$
 where ${\bar n_j} = n_j + \nu ( j-1)  $ are quasi-excitation numbers.  
However it is evident that these ${\bar n_j} $s 
   are no longer integers and they 
satisfy a modified selection rule given by 
  ${\bar n_{j+1}} - {\bar n_j}  \geq  \nu $,
 which restricts the difference
between the quasi-excitation numbers to be at least  
$ \nu$ apart.
As a consequence, the 
 Calogero model (\ref{1}) provides a microscopic 
realisation for fractional statistics with $\nu $ representing
the corresponding GES parameter \cite{is,mu,po,bk}.

Recently,  theoretical investigations on
different nonhermitian Hamiltonians have received a major
boost because many such systems, 
whenever they are invariant under combined parity 
and time reversal (PT) symmetry, lead to real energy 
eigenvalues \cite{pt1,pt2,pt3,pt4}.
This seems to suggest that the condition of hermiticity 
on a Hamiltonian can  be replaced by the weaker condition of PT symmetry 
to ensure that the corresponding eigenvalues would be real ones. 
However, till now this is merely a conjecture supported by several examples.
Moreover, in almost all of these examples, the Hamiltonians of 
only one particle in one space dimension have been considered.  
Therefore, it should be interesting  to test 
 this conjecture for the cases of nonhermitian $N$-particle Hamiltonians
 in one dimension which remain invariant
under the $PT$ transformation \cite{bm}:
\beq
i \rightarrow -i, \ \ x_j \rightarrow - x_j, \ \  p_j \rightarrow p_j \, ,
\label {4}
\eeq 
where $j \in [1,2, \cdots , N],$  and 
$x_j$ ($p_j \equiv -i \frac {\partial }{\partial  x_j}$) denotes
the coordinate (momentum) operator of the $j$-th particle. 
In particular, one may construct an extension of Calogero model by
adding to it some nonhermitian but PT invariant interaction,  
and enquire whether such extended model would lead to 
real spectrum.

The purpose of the present article is to review the progress \cite{bk,bm}
on the above mentioned problem for some special cases,
where the PT invariant extension of Calogero model can 
be solved exactly. In Sec.2 of this article we consider such a PT invariant 
extension of $A_{N-1}$ Calogero model and show that,
within a certain range of the related parameters, this 
extended Calogero model yields real eigenvalues. Next,
in Sec.3, we explore the connection of these real eigenvalues with 
fractional statistics. Section 4 is the concluding section.

\noindent \section { Exact solution of an extended Calogero model}
\renewcommand{\theequation}{2.{\arabic{equation}}}
\setcounter{equation}{0}

Let us consider a Hamiltonian of the form \cite{bk} 
\beq
 {\cal H} = H +  \delta \sum_{j\neq k} {1 \over x_j - x_k}{\d \over \d x_j},
\label {5}
\eeq 
where $H $ is given by eqn.(\ref {1}) and 
$\delta $ is a real parameter.
It may be observed that though the Hamiltonian 
(\ref{5}) violates hermiticity property due to the presence of momentum
dependent term  like 
$ \delta \sum_{j\neq k} {1 \over x_j - x_k}{\d \over \d x_j} $, it 
remains invariant under the combined PT transformation (\ref {4}).
Next we recall that, 
 $A_{N-1}$ and $B_N$ Calogero models as well as their distinguishable variants
have been solved recently by mapping them to a system of 
free oscillators  \cite{gp,so,bf,wa}.
With the aim of solving the PT invariant extension 
(\ref{5}) of $A_{N-1}$ Calogero model through a similar produce, 
we assume that (justification for this assumption will be given later)
the ground state wave function for this extended 
model is given by 
\beq
\psi_{gr}= e^{- {\omega\over 2}  \sum_{j=1}^N x_j^2 }
\prod_{j<k} (x_j-x_k)^\nu  , 
\label {6}
\eeq
where $\nu $ is a real positive number which is 
 related to the coupling constants 
$g$ and $\delta $ as 
\beq 
g = \nu^2 - \nu ( 1+ 2 \delta ) \, .
\label {7}
\eeq
Now if we use the expression 
 (\ref {6}) for a similarity transformation to 
 the Hamiltonian (\ref {5}), it reduces to an `effective Hamiltonian'  
 of the form
\beq
{\cal H}'=\psi_{gr}^{-1} {\cal H} \psi_{gr}  = S^- + \omega S^3 + E_{gr} \, ,
\label {8}
\eeq
where the Lassalle operator ($S^-$) and Euler operator ($S^3$) are given by
\beq 
 S^- = -  {1\over 2} \sum_{j=1}^N {\d^2 \over \d x_j^2} -
  (\nu- \delta)\sum_{j \neq k}
  {1 \over x_j - x_k}{\d \over \d x_j}, ~~~
  S^3 =  \sum_{j=1}^N  x_j { \d \over \d x_j } ,
\label {9}
\eeq
and 
\beq
E_{gr}= {N \omega \over 2} \left [ 1 + (N-1)(\nu -\delta ) \right ] .
\label {10}
\eeq
It is easy to see that 
 the Lassalle operator and Euler operator, as defined 
 in eqn.(\ref {9}), satisfy the simple 
commutation relation: 
$ [S^3,S^-]= -2 S^-$. Using therefore the well known Baker-Hausdorff 
transformation we can remove the $S^-$ part of the effective Hamiltonian
${\cal H}' $ and through some 
additional similarity transformations reduce it finally to the 
free oscillator model \cite{bk}
\beq
H_{free} = {\cal S}^{-1} \left({\cal H}'- E_{gr}\right) {\cal S}
  = -  {1\over 2} \sum_{j=1}^N {\d^2 \over \d x_j^2} +
 {\omega^2 \over 2}  \sum_{j=1}^N x_j^2 - {\omega N \over 2},
\label {11}
\eeq
where ${\cal S} =e^{ {1\over 2 \omega } S^- } 
e^{ {1\over 4 \omega } \nabla^2 } 
e^{ {\omega \over 2} \sum_{j=1}^N x_j^2 }  $ and 
$ \nabla^2 = \sum_{j=1}^N {\d^2 \over \d x_j^2} $.

Due to similarity transformations in (\ref {8}) and (\ref {11}), 
one may naively think that 
 the eigenfunctions of the extended Calogero model (\ref {5}) can be 
obtained from those of free oscillators as: 
$ \psi_{n_1, n_2 ,\cdots, n_N} = \psi_{gr}  {\cal S} \left\{ \prod_{j=1}^N 
e^{ - {\omega \over 2} x_j^2 } H_{n_j}(x_j) \right \}$, where 
$n_j$s are arbitrary non-negative integers and 
$H_{n_j}(x_j)$ denotes the Hermite polynomial of order $n_j$. 
However it is easy to check that,
similar to the case of $A_{N-1}$ Calogero model \cite{gp}, 
the action of ${\cal S}$ on free oscillator eigenfunctions 
leads to a singularity unless they are symmetrised with respect to 
all coordinates. Therefore, nonsingular eigenfunctions of 
  the extended Calogero model (\ref {5}) can be 
obtained from the 
eigenfunctions of free oscillators as
\beq 
 \psi_{n_1, n_2, \cdots , n_N} = \psi_{gr} \ {\cal S} \Lambda_+ 
\left\{ \prod_{j=1}^N 
e^{ - {\omega \over 2} x_j^2 } H_{n_j}(x_j) \right \} \, ,
\label {12}
\eeq
where $\Lambda_+ $ completely symmetrises all coordinates and thus 
projects the distinguishable many-particle wave functions to the 
bosonic part of the Hilbert space. Evidently, the eigenfunctions 
(\ref {12}) will be mutually independent if the excitation 
numbers $n_j$s obey the bosonic selection rule: $n_{j+1} \geq n_j $.
 Thus, in spite of the fact that the interacting Hamiltonian (\ref {5})
is convertible to the free oscillator model, the need for symmetrization 
shows that the many-particle correlation is in fact inherent in this model.  
The eigenvalues of the Hamiltonian (\ref{5})
 corresponding to the states  (\ref{12}) will  
  naturally be given by \cite{bk} 
\beq 
E_{n_1,n_2, \cdots , n_N} = E_{gr}+\omega \sum_{j=1}^N n_j= 
 {N \omega \over 2} \left [ 1 + (N-1)(\nu -\delta ) \right ] +
 \omega \sum_{j=1}^N n_j .
\label {13}
\eeq
Since $\delta $ and $\nu $ are 
real parameters, the energy eigenvalues 
(\ref {13}) are also real ones. Thus we interesting find that 
the nonhermitian PT invariant Hamiltonian (\ref {5}) yields a real 
spectrum. Furthermore, it is evident that for all $n_j=0,$  
the energy $E_{n_1,n_2, \cdots , n_N}$ 
  attains its minimum value   
$E_{gr}$. At the same time, as can be easily worked out
 from eqn.(\ref {12}),
 the corresponding  eigenfunction  reduces to 
$\psi_{gr}$ (\ref {6}). 
This proves that $\psi_{gr}$ is indeed the ground state wave function 
for Hamiltonian (\ref{5}) with eigenvalue $E_{gr}$. 

It may be observed that the eigenfunctions (\ref {12}) pick 
up a phase factor $(-1)^\nu $ under the exchange of any two 
particles. Therefore, 
 $\nu $ represents the exchange statistics parameter 
for the extended Calogero model (\ref {5}). By solving the quadratic
 eqn.(\ref {7}), one can explicitly write down $\nu $
 as a function of $g$ and $\delta $ as
\beq  \nu =  (\delta + {1\over 2}) \pm 
 \sqrt{ g + (\delta +{1\over 2})^2 } .
 \label {14} 
\eeq
For the purpose of obtaining real 
 eigenvalues (\ref {13}) as well as nonsingular eigenfunctions 
  (\ref {12})  at the limit $x_i 
\rightarrow x_j $,
we have assumed at the beginning of this section
that $\nu $ is a real positive parameter.  This assumption 
 leads to a restriction on the allowed values of the coupling 
constants $g$ and $\delta $ in the following way.
First of all,
for the case $g < - (\delta +{1\over 2})^2 $, 
 eqn.(\ref {14}) yields two imaginary solutions. 
Secondly, for the case $\delta < - {1\over 2} , ~~
0 > g >  - (\delta +{1\over 2})^2  $ 
 eqn.(\ref {14}) yields two real but negative solutions. 
Inequalities corresponding to these two cases represent two forbidden 
 regions of $( \delta ,g)$ plane which are excluded 
from our analysis.  

For the case $g>0$ with arbitrary value of $\delta $, one gets a real positive 
and a real negative solution from eqn.(\ref {14}). The real positive 
solution evidently leads to
   physically acceptable set of eigenfunctions and corresponding 
eigenvalues within this allowed region of $( \delta ,g)$ plane.
  Finally we consider the parameter range 
 $\delta > - {1\over 2} , ~~ 0 > g >  - (\delta +{1\over 2})^2  $,
for which eqn.(\ref {14}) yields two real positive solutions. 
It is easy to see that these 
 two real positive solutions are related to each other through a 
`reflection symmetry' given by $\nu \rightarrow 1 + 2\delta - \nu $.
Consequently, for each point on the 
 $( \delta , g )$ plane within this allowed parameter range, 
one obtains two different values of 
the exchange statistics parameter 
  leading to two distinct sets of physically acceptable 
eigenfunctions and eigenvalues.
Thus we curiously find that a kind of `phase transition' occurs 
at the line $\delta =
- {1 \over 2}$ on the $(\delta ,g)$ plane. 
For the case $\delta > - {1 \over 2}$,
 exchange statistics parameter 
shows the reflection symmetry
 when $g$ is chosen  within an interval 
$ - ( {1\over 2} + \delta )^2 < g \leq 0$. 
On the other hand 
for the case $\delta \leq - {1 \over 2}$, 
such reflection symmetry is lost for any possible value of $g$.

We have seen in this section that, 
similar to the case of $A_{N-1}$ Calogero model, 
the extended model (\ref {5}) can also be solved by mapping it 
to a system of free harmonic oscillators.
So it is natural to enquire whether 
this  extended  model is directly related to the 
 $A_{N-1}$ Calogero model through some similarity transformation.
Investigating along this line \cite{bm}, we find that 
\beq
\Gamma ^{-1}{\cal H} \Gamma =  H^\prime =
  \frac{ 1}{2}\sum p_j^2 + \frac{ 1}{2}\omega ^2
\sum x_j^2  + g^\prime 
\sum_{j\neq k}^N \frac{1}{(x_j-x_k)^2} \, ,
\label{15}
\eeq
where  $ \Gamma = \prod_{j< k} (x_j-x_k)^\delta  $,
and $H^\prime $ denotes the Hamiltonian of $A_{N-1}$ 
Calogero model with   `renormalised' coupling constant 
 given by $g^\prime = g + \delta (1+\delta ) $. Due to the existence of 
such similarity transformation, one may expect that the Hamiltonians 
  ${\cal H}$ and $H^\prime$ always lead to exactly same eigenvalues.
However it should be noted that,
  within a parameter range given by
$\delta > 0 , ~g > - \delta (1+ \delta )$, there exists 
a positive solution of eqn.(\ref {7}) satisfying the condition
$  \nu - \delta < 0 $. Therefore, we can not get any lower bound 
for the corresponding  energy eigenvalues  (\ref {13}) at $N
\rightarrow \infty $ limit. On the other hand, the 
  energy eigenvalues  (\ref {2}) of $A_{N-1}$ Calogero model are certainly
bounded from below for all possible choice of $N$ and $g$.
So there exists
a parameter range within which the spectrum of extended Calogero model differs 
qualitatively from the spectrum of the original Calogero model.
To explain this rather unexpected result, we first observe that 
the renormalised coupling constant 
$ g^\prime $ would be a positive quantity within the 
above mentioned parameter range.
Consequently, the corresponding exclusion statistics parameter 
$\nu^\prime $, which is obtained by solving eqn.(\ref {3}),
  has one positive  and one negative solution. 
 One usually throws away this negative solution of 
 $  \nu^\prime $, since the corresponding eigenfunctions 
 become singular at the limit $x_j \rightarrow x_k$.  However,
by using the relation (\ref {15}), such  singular eigenfunctions 
(denoted by $\psi^\prime(x_1,x_2, \cdots ,x_N$)) 
may now be used to construct the 
 eigenfunctions of extended Calogero model
(denoted by $\psi(x_1, x_2, \cdots x_N$)) as
\beq
\psi (x_1, x_2, \cdots , x_N ) ~=~ 
 \prod_{j < k} (x_j-x_k)^\delta 
  \psi^\prime (x_1, x_2, \cdots , x_N ) \,  .
\label{16}
\eeq
It can be easily checked that, due to the existence of the factor
 $\prod_{j < k} (x_j-x_k)^\delta $, the r.h.s. of the above equation 
 becomes nonsingular at the limit $x_j \rightarrow x_k \ $. 
   Thus we curiously find that 
singular eigenfunctions of $H^\prime $ can be used to generate  nonsingular 
  eigenfunctions of ${\cal H}$  through the relation (\ref {16}).  
This shows that the similarity transformation (\ref {15}) is a subtle one 
and, within a certain parameter range,  eigenvalues of extended
Calogero model  will match with those of Calogero model (having
renormalised coupling constant) only if 
the corresponding unphysical eigenfunctions are taken into account. 

\vspace{.75cm}
\noindent \section {Connection with fractional statistics}
\renewcommand{\theequation}{3.{\arabic{equation}}}
\setcounter{equation}{0}

We have mentioned in Sec.1 that GES  
can be realised microscopically in 
$A_{N-1}$ Calogero model with hermitian Hamiltonian.
The GES parameter for this Calogero model
 is a measure of `level repulsion' of the
quantum numbers generalising the Pauli exclusion principle.
Now for exploring GES in the case of PT invariant model (\ref {5}), 
we observe that  eqn.(\ref{13}) can be rewritten \cite{bk} 
exactly in the form of energy spectrum for $N$ free oscillators as
\beq
E_{n_1,n_2\cdots n_N} = \frac{N \omega }{2} + \omega
\sum_{j=1}^N \bar{n}_j \, ,
\label{17}
\eeq
where
\beq
\bar{n}_j = n_j +  (\nu - \delta ) (j-1) 
\label {18}
\eeq
are quasi-excitation numbers.
 However, from eqn.(\ref {18}) it is evident
that such quasi-excitation numbers are no longer integers
and satisfy a modified selection rule:
$ \bar{n}_{j+1}- \bar{n}_j \geq  \nu - \delta \,  .$  Since the minimum 
 difference between two consecutive 
$\bar{n}_j $s is given by
\beq
 \tilde{\nu } = \nu - \delta_ ,
\label {19}
\eeq
the spectrum of extended $A_{N-1}$ Calogero model (\ref {5})
satisfies GES with parameter $ \tilde{\nu}$ \cite{bk}.  
Several comments about this GES 
parameter are in order. 
It may be to noted that for $\delta \neq 0$, the GES parameter 
$\tilde \nu $ is different from the power 
index $ \nu $, which is responsible for the 
symmetry of the wave function. 
Therefore we may interestingly conclude that unlike  Calogero model,
 the exclusion statistics for  model
(\ref{5}) differs from its exchange statistics.  Furthermore it is already
 noticed that, on a region of
 $(\delta ,g)$ plane satisfying the inequalities
$\delta > 0 , ~g > - \delta (1+ \delta )$, there exists 
a positive solution of eqn.(\ref {7}) which yields a negative 
value of $\tilde \nu $. For this case, however,  one does not
 get well defined thermodynamic 
relations at $N\rightarrow \infty$ limit and, therefore, 
 can not interpret $\tilde \nu $ as the GES parameter.

By using eqn.(\ref {7})
and (\ref {19}), we find the relation 
\beq
  \tilde{\nu }^2  - \tilde{\nu }   = g + \delta ( \delta + 1) ,
\label {20}
\eeq
which clearly  describes a parabolic
curve  in the coupling constant plane $(\delta , g) $ for any fixed value 
of $\tilde { \nu} $.
As a consequence of this, the competing effect of the independent
coupling constants $g$ and $\delta $  can  make
   the GES feature of  (\ref{5}) much
  richer in comparison with the  Calogero model.
For example, while
bosonic (fermionic) excitations in the Calogero model
 occur only in the absence of long-range interaction,
 the quasi-excitations in  (\ref{5}) can behave as 
pure bosons (fermions)  
 even in the presence of both the long-range
  interactions satisfying the constraint
$\tilde \nu(\delta ,g)=0$ \ ($\tilde \nu(\delta ,g)=1$). 
Both of these constraints
 lead to 
the same  parabolic curve $g=-\delta (1+\delta )$. 
A family  of such parabolas with shifted  apex points are generated 
 for other values of 
$\tilde \nu$ and the lowest apex point is attained at 
$\tilde \nu= {1\over 2}$,
where
the quasi-excitations would behave as semions.

\noindent \section{Conclusion}

Here we construct a many-particle
quantum system (\ref {5}) by adding some nonhermitian 
but combined parity and time reversal (PT)
 invariant interaction to the $A_{N-1}$ Calogero model.
By using appropriate similarity transformations, we are able 
to map this extended Calogero model to a set of free harmonic oscillators
and solve this model exactly. It turns out that this 
many-particle system with nonhermitian Hamiltonian 
yields a real spectrum. This fact supports the conjecture that
  the condition of hermiticity 
on a Hamiltonian can be replaced by the weaker condition of PT symmetry 
to ensure that the corresponding  eigenvalues would be real ones. 
It is also found that the spectrum of extended 
 Calogero model obeys a selection rule which leads to generalised 
exclusion statistics (GES). 

However, this extended Calogero model exhibits some remarkable properties 
which are absent in the case of usual
  Calogero model. For example, we curiously find that the
 GES parameter for this extended Calogero model 
 differs from the corresponding exchange statistics parameter. Moreover
a `reflection symmetry' of the 
  exchange statistics parameter, which  is known to exist for 
   $A_{N-1}$ Calogero model, can  be found
 in the case of extended model only if  
 the strength of  PT invariant interaction exceeds a critical value. 

Finally we note that,
it is possible to obtain another exactly solvable many-particle
quantum system  by adding some nonhermitian 
but PT invariant interactions to the $B_N$ Calogero model (associated
with $B_N$ Lie algebra) \cite{bm}. Such a PT invariant 
  model also leads to real spectrum with 
properties quite similar to the case of
 extended $A_{N-1}$ Calogero model. 

\bigskip

\noindent {\bf Acknowledgments }

I would like to thank Prof. A. Kundu and Dr. B.P. Mandal for 
collaborating with me in Ref.7 and Ref.12. I also like to
thank Prof. M.L. Ge for kindly inviting me to the 
`APCTP-Nankai Symposium', October 7-10, 2001,  Tianjin, China.


\begin{thebibliography}{99}

\bibitem{ha}F. D. M. Haldane, {\em Phys. Rev. Lett.} {\bf 67} (1991) 937.

\bibitem{ca}F. Calagero, {\em J. Math. Phys.} {\bf 10} (1969) 2191; 
{\em J. Math. Phys.} {\bf 12} (1971) 419.

\bibitem{su}B. Sutherland, {\em J. Math. Phys.} {\bf 12} (1971) 246.

\bibitem{is}S.B. Isakov, {\em Mod. Phys. Lett. } {\bf B 8} (1994) 319; 
{\em Int. J. Mod. Phys. } {\bf A 9} (1994) 2563.

\bibitem{mu}M.V.N. Murthy and R. Shankar, {\em Phys. Rev. Lett.}
{\bf 73} (1994) 3331.

\bibitem{po} A.P. Polychronakos, in 
{\em Les Houches Lectures (Summer 1998)}, hep-th/9902157.

\bibitem{bk}B. Basu-Mallick and A. Kundu, 
{\em Phys. Rev. } {\bf B 62} (2000) 9927.    

\bibitem{pt1}C.M. Bender and S. Boettcher, {\em Phys. Rev. Lett.}
{\bf 80} (1998) 5243; {\em J. Phys. } {\bf A 31} (1998) L273.

\bibitem{pt2}C.M. Bender, S. Boettcher and P.N. Meisinger,
{\em J. Math. Phys.} {\bf 40} (1999) 2210.

\bibitem{pt3} F.M. Fernandez, R. Guardiola, J. Ros and M. Znojil, 
{\em J. Phys.  }{\bf A 32} (1999)  3105.

\bibitem{pt4} A. Khare and B. P. Mandal,
{\em Phys. Lett. } {\bf A 272} (2000) 53.

\bibitem{bm}B. Basu-Mallick and B.P. Mandal, 
{\em Phys. Lett. } {\bf A 284} (2001) 231.

\bibitem{gp}N. Gurappa and P.K. Panigrahi, 
{\em Phys. Rev.} {\bf B 59} (1999) R2490.

\bibitem{so}K. Sogo, {\em J. Phys. Soc. Jpn.} {\bf 65} (1996) 3097.

\bibitem{bf}T.H. Baker and P.J. Forrester, {\em Nucl. Phys.}
{\bf B 492} (1997) 682.

\bibitem{wa}H. Ujino, A. Nishino and M. Wadati, 
{\em J. Phys. Soc. Jpn.} {\bf 67} (1998) 2658.

\end{thebibliography}
\end{document}